\documentclass[runningheads]{llncs}
\usepackage[T1]{fontenc}
\usepackage[misc]{ifsym}
\usepackage{multirow}
\usepackage{makecell}
\usepackage{graphicx}
\usepackage{amsmath}
\usepackage{amssymb}
\usepackage{color}
\usepackage{xfrac}
\usepackage{wrapfig}
\usepackage{array}  

\usepackage{bm}

\def\mP{\mathcal{P}}

\begin{document}

\title{MEPNet: A Model-Driven Equivariant Proximal Network for Joint  Sparse-View Reconstruction and Metal Artifact Reduction in CT Images}

\titlerunning{Model-Driven Equivariant Proximal Network}

%\titlerunning{MEPNet}

\author{Hong Wang\textsuperscript{(\Letter)} \and
Minghao Zhou \and
Dong Wei \and
Yuexiang Li\and
Yefeng Zheng}
%1{Wang, Hong}
%2{Zhou, Minghao}
%3{Wei, Dong}
%4{Li, Yuexiang}
%5{Zheng, Yefeng}

\authorrunning{H. Wang et al.}
%\institute{Princeton University, Princeton NJ 08544, USA \and
%Springer Heidelberg, Tiergartenstr. 17, 69121 Heidelberg, Germany
\institute{Tencent Jarvis Lab, Shenzhen, P.R. China\\
\email{\{hazelhwang,hippomhzhou,donwei,vicyxli,yefengzheng\}@tencent.com} 
}

% \author{Paper ID 831}
% %\institute{Anonymous Organization\\ \email{**@******.***}}
% \institute{Anonymous Organization}
\maketitle              % typeset the header of the contribution
\begin{abstract}
Sparse-view computed tomography (CT) has been adopted as an important technique for speeding up data acquisition and decreasing radiation dose. However, due to the lack of sufficient projection data, the reconstructed CT images often present severe artifacts, which will be further amplified when patients carry metallic implants. For this joint sparse-view reconstruction and metal artifact reduction task, most of the existing methods are generally confronted with two main limitations: 1) They are almost built based on common network modules without fully embedding the physical imaging geometry constraint of this specific task into the dual-domain learning; 2) Some important prior knowledge is not deeply explored and sufficiently utilized. Against these issues, we specifically construct a dual-domain reconstruction model and propose a model-driven equivariant proximal network, called MEPNet. The main characteristics of MEPNet are: 1) It is optimization-inspired and has a clear working mechanism; 2) The involved proximal operator is modeled via a rotation equivariant convolutional neural network, which finely represents the inherent rotational prior underlying the CT scanning that the same organ can be imaged at different angles. Extensive experiments conducted on several datasets comprehensively substantiate that compared with the conventional convolution-based proximal network, such a rotation equivariance mechanism enables our proposed method to achieve better reconstruction performance with fewer network parameters. \textit{We will release the code at \url{https://github.com/hongwang01/MEPNet}.}
\end{abstract}

\section{Introduction}
Computed tomography (CT) has been widely adopted in clinical applications. To reduce the radiation dose and shorten scanning time, sparse-view CT has drawn much attention in the community~\cite{zhang2018sparse,lee2018deep}. However, sparse data sampling inevitably degenerates the quality of CT images and leads to adverse artifacts. In addition, when patients carry metallic implants, such as hip prostheses and spinal implants~\cite{liao2019adn,wang2021indudonet,liu2020deep}, the artifacts will be further aggravated due to beam hardening and photon starvation. For the joint sparse-view reconstruction and metal artifact reduction task (SVMAR), how to design an effective method for artifact removal and detail recovery is worthy of in-depth exploration.

For the sparse-view (SV) reconstruction, the existing deep-learning (DL)-based methods can be roughly divided into three categories based on the information domain exploited, \emph{e.g.}, sinogram domain, image domain, and dual domains. Specifically, for the sinogram-domain methods, sparse-view sinograms are firstly repaired based on deep networks, such as U-Net~\cite{lee2018deep} and dense spatial-channel attention network~\cite{zhou2021limited}, and then artifact-reduced CT images are reconstructed via the filtered-back-projection (FBP) process. For the image-domain methods, researchers have proposed to learn the clean CT images from degraded ones via various structures~\cite{zhang2018sparse,zhang2021transct,shen2022nerp}. Alternatively, both sinogram and CT images are jointly exploited for the dual reconstruction~\cite{zhang2020metainv,wang2021dudotrans,zhou2022dudodr,ding2021learnable}.

% To fully exploit the informative sinograms, the dual-domain methods~\cite{zhang2020metainv,wang2021dudotrans,zhou2022dudodr,ding2021learnable} propose to jointly reconstruct the sinogram and CT image. 

For the metal artifact reduction (MAR) task, similarly, the current DL-based approaches can also be categorized into three types. To be specific, sinogram-domain methods aim to correct the sinogram for the subsequent CT image reconstruction~\cite{zhang2018convolutional,ghani2019fast}.
Image-domain-based works have proposed different frameworks, such as simple residual network~\cite{huang2018metal} and an interpretable structure~\cite{wang2021dicdnet}, to learn artifact-reduced images from metal-affected ones. The dual-domain methods~\cite{lin2019dudonet,wang2021indudonet,zhou2022dudodr} focus on the mutual learning between sinogram and CT image.

Albeit achieving promising performance, these aforementioned methods are sub-optimal for the SVMAR task. The main reasons are: 1) Most of them do not consider the joint influence of sparse data sampling and MAR, and do not fully embed the physical imaging constraint between the sinogram domain and CT image domain under the SVMAR scenario; 2) Although a few works focus on the joint SVMAR task, such as~\cite{zhou2022dudodr}, the network structure is empirically built based on off-the-shelf modules, \emph{e.g.}, U-Net and gated recurrent units, and it does not fully investigate and embed some important prior information underlying the CT imaging procedure. However, for such a highly ill-posed restoration problem, the introduction of the proper prior is important and valuable for constraining the network learning and helping it evolve in a right direction~\cite{wang2021indudonet}.

To alleviate these issues, in this paper, we propose a model-driven equivariant proximal network, called MEPNet, which is naturally constructed based on the CT imaging geometry constraint for this specific SVMAR task, and takes into account the inherent prior structure underlying the CT scanning procedure. Concretely, we first propose a dual-domain reconstruction model and then correspondingly construct an unrolling network framework based on a derived optimization algorithm. Furthermore, motivated by the fact that the same organ can be imaged at different angles making the reconstruction task equivariant  to rotation~\cite{chen2021equivariant}, we carefully formulate the proximal operator of the built unrolling neural network as a rotation-equivariant convolutional neural network (CNN). Compared with the standard-CNN-based proximal network with only translation-equivariance property~\cite{cohen2016group}, our proposed method effectively encodes more prior knowledge, \emph{e.g.}, rotation equivariance, possessed by this specific task. With such more accurate regularization, our proposed MEPNet can achieve higher fidelity of anatomical structures and has better generalization capability with fewer network parameters. This is finely verified by comprehensive experiments on several datasets of different body sites. To the best of our knowledge, we should be the first to study rotation equivariance in the context of SVMAR and validate its utility, which is expected to make insightful impacts on the community.

% \section{Joint Spatial and Radon Domain Reconstruction model for Metal Artifact Reduction}
\begin{figure*}[t]
  \begin{center}
  %\vspace{-3mm}
     \includegraphics[width=1\linewidth]{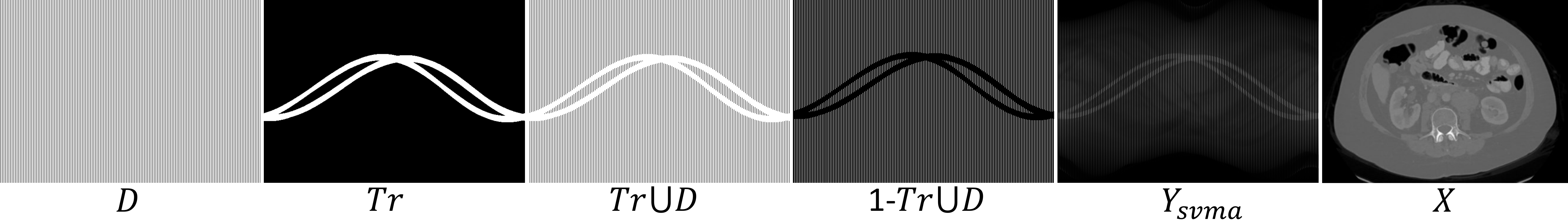}
  \end{center}
  \vspace{-6mm}
     \caption{Illustration of the elements in the model Eq.~\eqref{eq:model} for easy understanding.}
       \vspace{-5mm}
  \label{figmodel}
\end{figure*}

\vspace{-3mm}
\section{Preliminary Knowledge about Equivariance}
% From~\cite{cohen2016group,xie2022fourier}, equivariance of a mapping means that given $G$ as a group of transformations, we say that $\Phi$ is equivariant about the action of $G$ if transforming an input $f$ by a transformation $g$ and then feeding $T_{g}\big( f\big)$ into the mapping $\Phi$ obtains the same result as first passing $f$ through $\Phi$ and then transforming the representation $\Phi\big(f\big)$. Mathematically, 

Equivariance of a mapping w.r.t. a certain transformation indicates that executing the transformation on the input produces a corresponding transformation on the output~\cite{cohen2016group,xie2022fourier}. Mathematically, given a group of transformations $G$, a mapping $\Phi$ from the input feature space to the output feature space is said to be group equivariant about $G$ if
\vspace{-2mm}
\begin{equation}\label{eq:concept}
\Phi\big(T_{g}\big( f\big)\big) = T^{'}_{g}\big(\Phi\big(f\big)\big),  \forall g \in G,
\vspace{-2mm}
\end{equation}
where $f$ is any input feature map in the input feature space; $T_{g}$ and $T^{'}_{g}$ represent the actions of $g$ on the input and output, respectively.
% where $T_{g}$ and $T^{'}_{g}$ represent the actions of $g$ on the input and out, respectively.

The prior work~\cite{cohen2016group} has shown that adopting group equivariant CNNs to encode symmetries into networks would bring data efficiency and it can constrain the network learning for better generalization. For example, compared with the fully-connected layer, the translational equivariance property enforces weight sharing for the conventional CNN, which makes CNN use fewer parameters to preserve the representation capacity and then obtain better generalization ability. Recently, different types of equivariant CNNs have been designed to preserve more symmetries beyond current CNNs, such as rotation symmetry~\cite{chen2021equivariant,weiler2018learning,celledoni2021equivariant} and scale symmetry~\cite{gunel2022scale,sosnovik2019scale}. However, most of these methods do not consider specific designs for the SVMAR reconstruction. In this paper, we aim to build a physics-driven network for the SVMAR task in which rotation equivariance is encoded.

\vspace{-3mm}
\section{Dual-Domain Reconstruction Model for SVMAR}
In this section, for the SVMAR task, we derive the corresponding dual domain reconstruction model and give an iterative algorithm for solving it.
% In this section, we introduce the proposed joint spatial and Radon domain reconstruction model in details.

\noindent{\bf Dual-Domain Reconstruction Model.}
Given the captured sparse-view metal-affected sinogram {{$Y_{svma}\in \mathbb{R}^{N_{b}\times N_{p}}$}}, where $N_{b}$ and $N_{p}$ are the number of detector bins and projection views, respectively, to guarantee the data consistency between the reconstructed clean CT image $X\in \mathbb{R}^{H\times W}$ and the observed sinogram $Y_{svma}$, we can formulate the corresponding optimization model as~\cite{zhou2022dudodr}:
\vspace{-2mm}
\begin{equation}\label{eq:model}
    \min_{{X}}\left\|(1-Tr \cup  D )\odot (\mP X-Y_{svma})\right\|_{F}^{2} + \mu R(X),
    \vspace{-2mm}
\end{equation}
where $D\in \mathbb{R}^{N_{b}\times N_{p}}$ is the binary sparse downsampling matrix with 1 indicating the missing region; $Tr\in \mathbb{R}^{N_{b}\times N_{p}}$ is the binary metal trace with 1 indicating the metal-affected region; $\mP$ is forward projection; $R(\cdot)$ is a regularization function for capturing the prior of $X$; $\cup $ is the union set;  $\odot$ is the point-wise multiplication; $H$ and $W$ are the height and width of CT images, respectively; $\mu$ is a trade-off parameter. One can refer to Fig.~\ref{figmodel} for easy understanding.

To jointly reconstruct sinogram and CT image, we introduce the dual regularizers $R_{1}(\cdot)$ and $R_{2}(\cdot)$, and further derive Eq.~\eqref{eq:model} as:
\vspace{-1mm}
\begin{equation}\label{o2}
    \min_{{S,X}}\left\|\mP X-S\right\|_{F}^{2} +\lambda \left\|(1-Tr \cup D)\odot (S-Y_{svma})\right\|_{F}^{2}+ \mu_{1} R_{1}(S)\!+\mu_{2} R_{2}(X),
     \vspace{-1mm}
\end{equation}
where $S$ is the to-be-estimated clean sinogram; $\lambda$ is a weight factor. Following~\cite{wang2021indudonet}, we rewrite $S$ as $\bar{Y}\odot\bar{S}$ for stable learning, where $\bar{Y}$ and $\bar{S}$ are the normalization coefficient implemented via the forward projection of a prior image, and the normalized sinogram, respectively. Then we can get the final dual-domain reconstruction model for this specific SVMAR task as:
% where $\bar{Y}$ is normalization coefficient, usually set as the FP of a prior image $\bar{X}$, \emph{i.e.,} $\bar{Y} = \mP \bar{X}$;
\vspace{-1mm}
\begin{equation}\label{eq:recon}
\min_{{\bar{S}, X}}\left\|\mP X\!\!-\!\!\bar{Y}\!\odot\!\bar{S}\right\|_{F}^{2} \!+\!\lambda \left\|(1\!-\!Tr \!\cup \! D)\!\odot\! (\bar{Y}\!\odot\!\bar{S}\!\!-\!\!Y_{svma})\right\|_{F}^{2}\!+\! \mu_{1}\! R_{1}(\bar{S})\!+\!\mu_{2} R_{2}(X).
\vspace{-1mm}
\end{equation}
As observed, given $Y_{svma}$, we need to jointly estimate $\bar{S}$ and $X$. For $R_{1}(\cdot)$ and $R_{2}(\cdot)$, the design details are presented below.

% we adopt CNN to automatically learn the prior information contained in $\bar{S}$ and $X$ from training dataset. This would be more flexible than traditional manually-set manners~\cite{wang2021indudonet}. 
 \vspace{1mm}
\noindent {\bf Iterative Optimization Algorithm.} To solve the model~\eqref{eq:recon}, we utilize the classical proximal gradient technique~\cite{parikh2014proximal} to alternatively update the variables $\bar{S}$ and $X$. At iterative stage $k$, we can get the corresponding iterative rules:
\small
\begin{equation}\label{eq:rules}
\begin{split}
\hspace{-4mm}\bar{S}_{k}&\!=\! \mbox{prox}_{\mu_{1}\eta_{1}}\!\big(\!\bar{S}_{k-1} \!\!-\!\!\eta_{1}\!\big(\bar{Y}\!\!\odot\!\!\big(\bar{Y}\!\!\odot\!\!\bar{S}_{k-1}\!\!-\!\!\mP X_{k-1}\big) \!\!+\!\!\lambda\big(1\!\!-\!\!Tr \!\cup \!D\big)\!\odot\!\bar{Y}\!\!\odot\!\big(\bar{Y}\!\!\odot\!\!\bar{S}_{k-1}\!-\!Y\big) \big)\big), \\
\hspace{-4mm} X_{k} &=  \mbox{prox}_{\mu_{2}\eta_{2}}\big(X_{k-1}- \eta_{2}\mP^{T}\big(\mP X_{k-1}-\bar{Y}\odot\bar{S}_{k}\big)\big),
\end{split}
\end{equation}
\normalsize
where $\eta_{i}$ is stepsize; $\mbox{prox}_{\mu_{i}\eta_{i}}(\cdot)$ is proximal operator, which relies on the regularization term $R_{i}(\cdot)$. For any variable, its iterative rule in Eq.~\eqref{eq:rules} consists of two steps: an explicit gradient step to ensure data consistency and an implicit proximal computation $\mbox{prox}_{\mu_{i}\eta_{i}}(\cdot)$  which enforces the prior $R_{i}(\cdot)$ on the to-be-estimated variable. Traditionally, the prior form $R_{i}(\cdot)$ is empirically designed, \emph{e.g.}, $l_{1}$ penalty, which may not always hold in real complicated scenarios. Due to the high representation capability, CNN has been adopted to adaptively learn the proximal step in a data-driven manner for various tasks~\cite{yang2017admm,wang2021indudonet}. Motivated by their successes, in the next section, we will deeply explore the prior of this specific SVMAR task and carefully construct the network for $\mbox{prox}_{\mu_{i}\eta_{i}}(\cdot)$.

 \vspace{-4mm}
\section{Equivariant Proximal Network for SVMAR}\label{sec:details}
By unfolding the iterative rules~\eqref{eq:rules} for $K$ iterations, we can easily build the unrolling neural network. Specifically, at iteration $k$, the network structure is sequentially composed of:
% \small
% \begin{equation}\label{eq:net}
% \begin{split}
%         \bar{S}_{t}=  &\text{proxNet}_{\theta_{\bar{s}}^{(t)}}\big(\bar{S}_{t-1} -\eta_{1}\big(\bar{Y}\odot\big(\bar{Y}\odot\bar{S}_{t-1}-\mP X_{t-1}\big)\right.\right.\\ &\left.\left.+\lambda\big(1-Tr \cup D\big)\odot\bar{Y}\odot\big(\bar{Y}\odot\bar{S}_{t-1}-Y\big)  \big)\big), \\
% X_{t} =  & \text{proxNet}_{{\theta_{x}^{(t)}}}\big(X_{t-1}- \eta_{2}\mP^{T}\big(\mP X_{t-1}-\bar{Y}\odot\bar{S}_{t}\big)\big),
% \end{split}
% \end{equation}
% \normalsize
\small
\begin{equation}\label{eq:net}
\begin{split}
\hspace{-4mm}\bar{S}_{k}&\!=\!\text{proxNet}_{\theta_{\bar{s}}^{(k)}}\!\big(\!\bar{S}_{k-1} \!\!-\!\!\eta_{1}\!\big(\bar{Y}\!\!\odot\!\!\big(\bar{Y}\!\!\odot\!\!\bar{S}_{k-1}\!\!-\!\!\mP X_{k-1}\big) \!\!+\!\!\lambda\big(1\!\!-\!\!Tr \!\cup \!D\big)\!\odot\!\bar{Y}\!\!\odot\!\!\big(\bar{Y}\!\!\odot\!\!\bar{S}_{k-1}\!-\!Y\big)  \big)\big), \\
\hspace{-4mm} X_{k} &= \text{proxNet}_{{\theta_{x}^{(k)}}}\big(X_{k-1}- \eta_{2}\mP^{T}\big(\mP X_{k-1}-\bar{Y}\odot\bar{S}_{k}\big)\big),
\end{split}
\vspace{-5mm}
\end{equation}
\normalsize
where $\text{proxNet}_{\theta_{\bar{s}}^{(k)}}$ and $\text{proxNet}_{{\theta_{x}^{(k)}}}$ are proximal networks with parameters $\theta_{\bar{s}}^{(k)}$ and $\theta_{x}^{(k)}$ to execute the proximal operators $\mbox{prox}_{\mu_{1}\eta_{1}}$ and $\mbox{prox}_{\mu_{2}\eta_{2}}$, respectively.

To build $\text{proxNet}_{\theta_{\bar{s}}^{(k)}}$, we follow~\cite{wang2021indudonet} and choose a standard-CNN-based residual structure with four [{\small{\emph{Conv+BN+ReLU+Conv+BN+Skip Connection}}}] residual blocks. While for $\text{proxNet}_{{\theta_{x}^{(k)}}}$, we carefully investigate that during the CT scanning, the same body organ can be imaged at different rotation angles. However, the conventional CNN for modeling $\text{proxNet}_{{\theta_{x}^{(k)}}}$ in~\cite{wang2021indudonet} has only the translation equivariance property and it cannot preserve such an intrinsic rotation equivariance structure~\cite{cohen2016group}. To tackle this problem, we propose to replace the standard CNN in~\cite{wang2021indudonet} with a rotation equivariant CNN. In this manner, we can embed more useful prior, such as rotation equivariance, to constrain the network, which would further boost the quality of reconstructed CT images (refer to Sec.~\ref{sec:exps}). 

Specifically, from Eq.~\eqref{eq:concept}, for a rotation group $G$ and any input feature map $f$, we expect to find a properly parameterized convolutional filter $\psi$ which is group equivariant about $G$, satisfying
% a convolutional filter $\psi$ is equivariant to the rotation transformation $T_{r}$ if for any input feature map $f$,
\begin{equation} \label{eq:7}
[T_{\theta}[f]] \star \psi  = T_{\theta}[f \star \psi] = f \star \pi_\theta[\psi]  , \forall \theta \in G,
\end{equation}
where $\pi_{\theta}$ is a rotation operator. Due to its solid theoretical foundation, the Fourier-series-expansion-based method~\cite{xie2022fourier} is adopted 
% In order to enforce rotation equivariance, 
to parameterize $\psi$ as:
\vspace{-1mm}
\begin{equation}\label{eq:filter}
    \psi (x) = \sum_{m=0}^{p-1} \sum_{n=0}^{p-1} a_{mn} \varphi_{mn}^{c}\big(x\big) +b_{mn}\varphi_{mn}^{s}\big(x\big),
    \vspace{-1mm}
\end{equation}
where $x=[x_{i},x_{j}]^{T}$ is 2D spatial coordinates; $a_{mn}$ and $b_{mn}$ are learnable expansion coefficients; $\varphi_{mn}^{c}\big(x\big)$ and $\varphi_{mn}^{s}\big(x\big)$ are 2D fixed basis functions as designed in~\cite{xie2022fourier}; $p$ is chosen to be 5 in experiments.
% $a_{mn}$ and $b_{mn}$ are learnable expansion coefficients; $\varphi_{mn}^{c}\big(x\big)$ and $\varphi_{mn}^{s}\big(x\big)$ are 2D fixed basis functions as:\footnote{Please refer to~\cite{xie2022fourier} for more explanations.}
% \vspace{-2mm}
% \small
% \begin{equation}\label{eq3}
% \varphi_{mn}^{c}\!\big(x\big)\!\!=\!\!\Omega \big(x\big)\!\text{cos}\!\big(\!\frac{2\pi}{ph}\!\left[\mathcal{I}_{p}\big(m\big),\mathcal{I}_{p}\big(n\big)\right]
%  \begin{bmatrix}  
%  x_{i}\\
% x_{j}
% \end{bmatrix}\!\big),\varphi_{mn}^{s}\!\big(x\big)\!\!=\!\!\Omega \big(x\big)\!\text{sin}\!\big(\!\frac{2\pi}{ph}\left[\mathcal{I}_{p}\big(m\big),\mathcal{I}_{p}\big(n\big)\right]
%  \begin{bmatrix}
%  x_{i}\\
% x_{j}
% \end{bmatrix}\!\big),
% \vspace{-2mm}
% \end{equation}
% \normalsize
% where the mesh size $h$ is set to $\frac{1}{4}$. 
% The rotation operator $T_{r_\theta}$ acted on $\psi$ can be achieved by coordinate transformation as:
The action $\pi_\theta$ on $\psi$ in Eq. (\ref{eq:7}) can be achieved by coordinate transformation as:
\vspace{-1mm}
\begin{equation}
\pi_\theta[\psi]\big(x\big) = \psi\big(U^{-1}_{\theta}x\big), \text{where}~U_{\theta} = \begin{bmatrix}
 \text{cos}\theta  & \text{sin}\theta\\
 \text{-sin}\theta  &\text{cos}\theta
\end{bmatrix}, \forall \theta \in G.
\vspace{-1mm}
\end{equation}
Based on the parameterized filter in Eq.~\eqref{eq:filter}, we follow~\cite{xie2022fourier} to implement the rotation-equivariant convolution for the discrete domain. Compared with other types, \emph{e.g.}, harmonics and partial-differential-operator-like bases~\cite{weiler2018learning,shen2020pdo}, the basis in Eq.~\eqref{eq:filter} has higher representation accuracy, especially when being rotated.

% This says that the correlation of a rotated image $T_{r}[f]$ with a filter $\psi$ is the same as the rotation of the original image $f$ convolved with the filter $\psi$. 
By implementing $\text{proxNet}_{\theta_{\bar{s}}^{(k)}}$ and $\text{proxNet}_{{\theta_{x}^{(k)}}}$ in Eq.~\eqref{eq:net} with the standard CNN and the rotation-equivariant CNN with the $p8$ group,\footnote{The parameterized filters for eight different rotation orientations share a set of expansion coefficients, largely reducing the network parameters (validated in Sec.~\ref{sec:exps}).} respectively, we can then construct the unrolling model-driven equivariant proximal network, called MEPNet. The expansion coefficients, $\{\theta_{\bar{s}}^{(k)}\}_{k=1}^{K}$, $\theta_{prior}$ for learning $\bar{Y}$~\cite{wang2021indudonet}, $\eta_{1}$, $\eta_{2}$, and $\lambda$, are all flexibly learned from training data in an end-to-end manner.
% Motivated by the recent studies~\cite{yang2017admm,yang2018proximal}, we propose a deep unfolding framework, namely InDuDoNet, for the MAR task.

%proposed by Yu \emph{et al.}
% U-Net~\cite{ronneberger2015u} is adopted as the backbone for Prior-net.
% we simply utilize the U-Net~\cite{ronneberger2015u} to learn $\bar{X}$ by taking the concatenation of metal-affected image $X_{ma}$ and LI corrected image $X_{LI}$ as input, where $X_{ma}$ and $X_{LI}$ are reconstructed from original metal-corrupted sinogram $Y$ and the linear interpolated sinogram $Y_{LI}$~\cite{kalender1987reduction}. Here the adopted U-Net has the similar structure as PriorNet in~\cite{yu2020deep} with the depth of 4,  but we halve the channel number for fewer network parameters.

%, together with the learning of $\bar{Y}$.
% We now introduce the network architecture for each stage. 
% One can refer to Fig.~\ref{fignet}(b) for better understanding. 
% As shown in Fig.~\ref{fignet} (b), 
% Each stage yields the sequential updates of $\bar{S}$ and $X$ generated by $\bar{S}$-net and $X$-net, respectively. 

Compared with most deep SV/MAR methods as well as~\cite{wang2021indudonet}, our proposed MEPNet has specific merits: 1) It is specifically constructed based on the physical imaging procedure for the SVMAR task, leading to a clear working mechanism; 2) It embeds more prior knowledge, \emph{e.g.}, rotation equivariance, which promotes better reconstruction; 3) It is desirable that the usage of more transformation symmetries would further decrease the number of model parameters and improve the generalization. These advantages are validated in the experiments below.

% first computed based on Eq.~(\ref{rulex}) and then sent into a deep network $\text{proxNet}_{\theta_{x}^{(n)}}(\cdot)$ with parameter $\theta_{x}^{(n)}$ to perform the mapping $\mbox{prox}_{\lambda_{2}\eta_{2}}(\cdot)$. Then, the updated artifact-reduced image is: $X_{n} = \text{proxNet}_{\theta_{x}^{(n)}}\big(\widehat{X}_{n-1}\big)$.
% {\itshape $\bar{S}$-net and $X$-net.}
% with the channel number as 32 and convolutional kernel size as $3\times3$.

% \noindent\textbf{Training Loss.} For a fair comparison, the training loss is set the same as~\cite{wang2021indudonet}:
% \small
% \vspace{-2mm}
% \begin{equation}\label{Loss}
% %\vspace{-1mm}
%   \mathcal{L} = \sum_{k=0}^{K}\alpha_{k}\left\|X_k-X_{gt} \right\|_F^2\odot(1-M)+w\big(\sum_{k=1}^{K}\alpha_{k}\left\| \bar{Y}\odot\bar{S}_k - Y_{gt}\right\|_F^2\big),
% \vspace{-1mm}
% \end{equation}
% \normalsize
% where $M$ is the binary metal mask; $X_{gt}$ and $Y_{gt}$ are clean image and sinogram, respectively; $w=0.1$; $\alpha_{k}=0.1$ ($k=0,\cdots, K-1$); $\alpha_{K}=1$~\cite{wang2021indudonet}.
%\textbf{Differences from Existing Iterative Methods}
% \vspace{-3mm}
\begin{figure*}[t]
  \begin{center}
  %\vspace{-3mm}
     \includegraphics[width=0.96\linewidth]{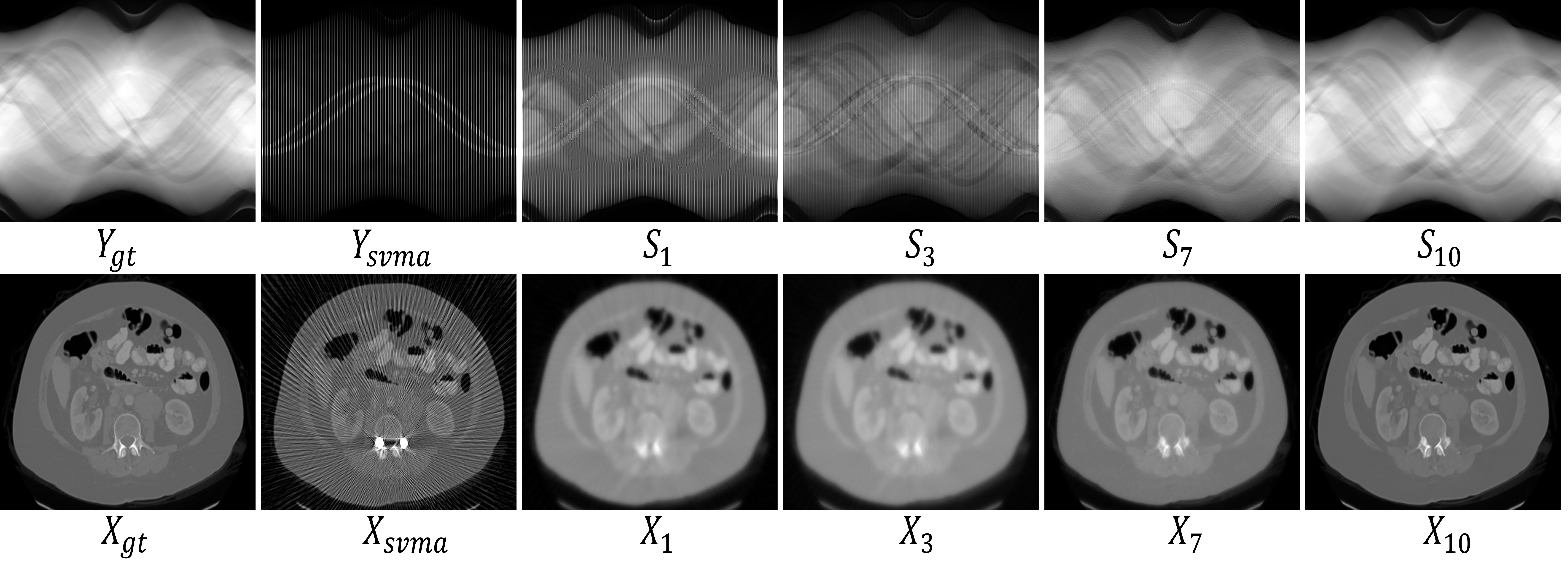}
  \end{center}
  \vspace{-7mm}
     \caption{The sinogram $S_{k}$ and CT image $X_{k}$ reconstructed by our MEPNet ($K$=10).}
  \label{fig:modelverif}
    \vspace{-4mm}
\end{figure*}
\vspace{-0mm}

\section{Experiments}

\subsection{Details Description}

\noindent\textbf{Datasets and Metrics.}
Consistent with~\cite{wang2021indudonet}, we synthesize the training set by randomly selecting 1000 clean CT images from the public DeepLesion~\cite{yan2018deep} and collecting 90 metals with various sizes and shapes from~\cite{zhang2018convolutional}. Specifically,
following the CT imaging procedure with fan-beam geometry in~\cite{yu2020deep,wang2021indudonet,zhou2022dudodr}, all the CT images are resized as $416\times 416$ pixels and 640 fully-sampled projection views are uniformly spaced in 360 degrees. To synthesize sparse-view metal-affected sinogram $Y_{svma}$, similar to~\cite{zhou2022dudodr}, we uniformly sample 80, 160, and 320 projection views to mimic 8, 4, and 2-fold radiation dose reduction. By executing the FBP process on $Y_{svma}$, we can obtain the degraded CT image $X_{svma}$.

The proposed method is tested on three datasets including DeepLesion-test (2000 pairs), Pancreas-test (50 pairs), and CLINIC-test (3397 pairs). Specifically, DeepLesion-test is generated by pairing another 200 clean CT images from DeepLesion~\cite{yan2018deep} with 10 extra testing metals from~\cite{zhang2018convolutional}. Pancreas-test is formed by randomly choosing 5 patients with 50 slices from Pancreas CT~\cite{roth2015deeporgan} and pairing each slice with one randomly-selected testing metal. CLINIC-test is synthesized by pairing 10 volumes with 3397 slices randomly chosen from CLINIC~\cite{liu2020deep} with one testing metal slice-by-slice. 
The 10 testing metals have different sizes as [2061, 890, 881, 451, 254, 124, 118, 112, 53, 35] in pixels. For evaluation on different sizes of metals as listed in Table~\ref{tab:160} below, we merge the adjacent two sizes into one group. Following~\cite{lin2019dudonet,wang2021indudonet}, we adopt peak signal-to-noise ratio (PSNR) and structured similarity index (SSIM) for quantitative analysis.

\noindent\textbf{Implementation Details.}
 Our MEPNet is trained end-to-end with a batch size of 1 for 100 epochs based on PyTorch~\cite{paszke2017automatic} on an NVIDIA Tesla V100-SMX2 GPU card. An Adam optimizer with parameters ($\beta_{1}$, $\beta_{2}$)=(0.5, 0.999) is exploited. The initial learning rate is $2\times10^{-4}$ and it is decayed by 0.5 every 40 epochs. For a fair comparison, we adopt the same loss function as~\cite{wang2021indudonet} and also select the total number of iterations $K$ as 10.

\vspace{-1mm}
\begin{figure*}[t]
  \begin{center}
  \vspace{-3mm}
     \includegraphics[width=0.95\linewidth]{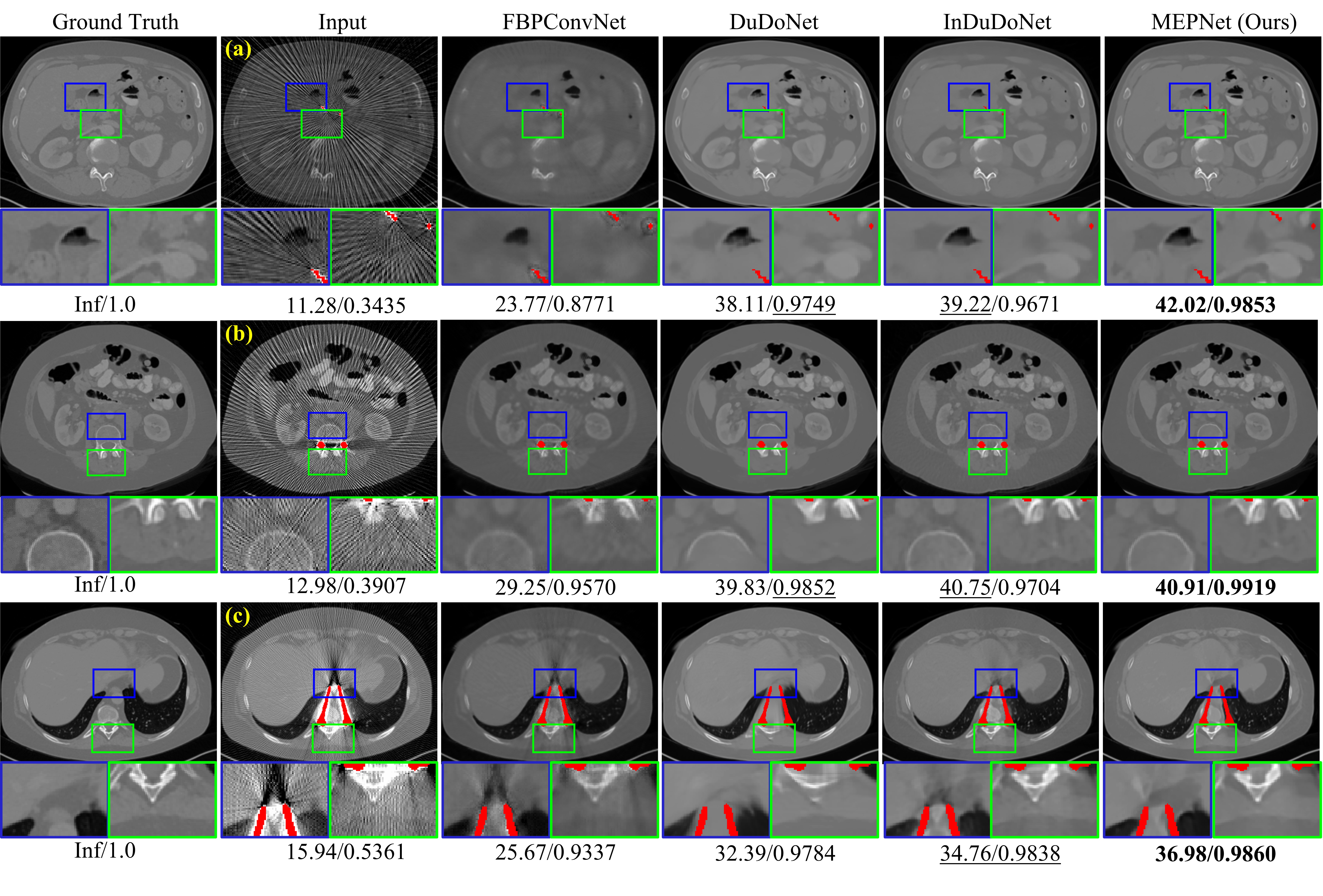}
  \end{center}
  \vspace{-9mm}
     \caption{DeepLesion-test: Artifact-reduced images (the corresponding PSNRs/SSIMs are shown below) of the comparing methods under different sparse-view under-sampling rates (a) ×8, (b) ×4, (c) ×2, and various sizes of metals marked by red pixels.}
  \label{fig:dp}
  \vspace{-1mm}
\end{figure*}
\begin{figure*}[t]
  \begin{center}
  %\vspace{-3mm}
     \includegraphics[width=0.95\linewidth]{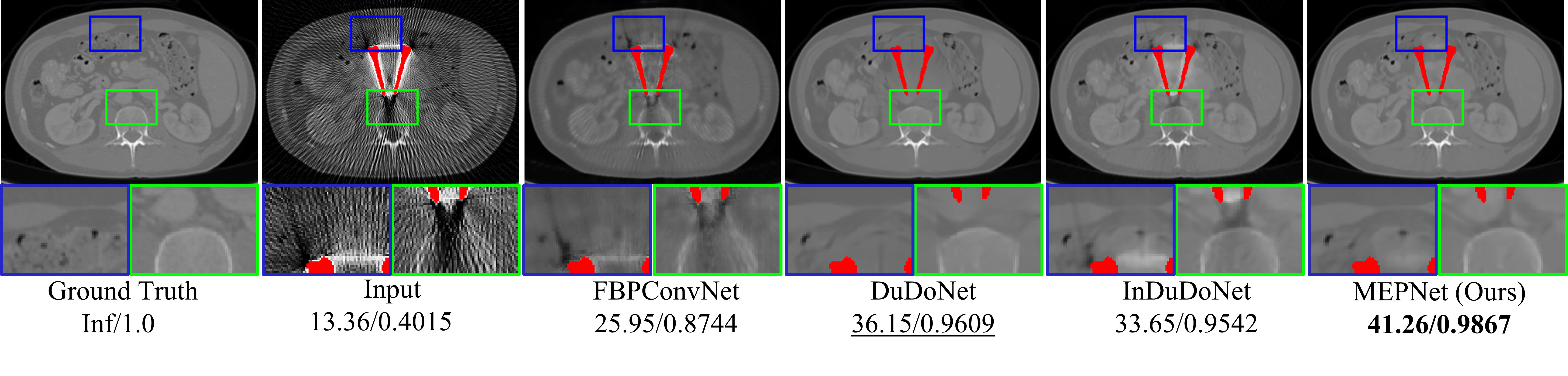}
  \end{center}
  \vspace{-9mm}
     \caption{Cross-domain: Reconstruction results with the corresponding PSNR/SSIM of different methods on Pancreas-test under ×4 under-sampling rate.}
  \label{fig:pan}
   \vspace{-4mm}
\end{figure*}
\subsection{Performance Evaluation}\label{sec:exps}
% \vspace{-2mm}
\noindent{\bf Working Mechanism.}
Fig.~\ref{fig:modelverif} presents the sinogram $S_{k}$ and CT image $X_{k}$ reconstructed by MEPNet at different stages. We can easily observe that $S_{k}$ and $X_{k}$ are indeed alternatively optimized in information restoration and artifact reduction, approaching the ground truth $Y_{gt}$ and $X_{gt}$, respectively. This finely shows a clear working mechanism of our proposed MEPNet, which evolves in the right direction specified by Eq.~(\ref{eq:rules}).

\noindent{\bf Visual Comparison.}
Fig.~\ref{fig:dp} shows the reconstructed results of different methods, including FBPConvNet~\cite{jin2017deep}, DuDoNet~\cite{lin2019dudonet}, InDuDoNet~\cite{wang2021indudonet}, and the proposed MEPNet, on three degraded images from DeepLesion-test with different sparse-view under-sampling rates and various sizes of metallic implants.\footnote{Here InDuDoNet is a particularly strong baseline and it is exactly an ablation study.} As seen, compared with these baselines, our proposed MEPNet can consistently produce cleaner outputs with stronger artifact removal and higher structural fidelity, especially around the metals, 
thus leading to higher PSNR/SSIM values. 

Fig.~\ref{fig:pan} presents the cross-domain results on Pancreas-test with ×4 under-sampling rate where DL-based methods are trained on synthesized DeepLesion. As seen, DuDoNet produces over-smoothed output due to the lack of physical geometry constraint on the final result. In contrast, MEPNet achieves more efficient artifact suppression and sharper detail preservation. Such favorable generalization ability is mainly brought by the dual-domain joint regularization and the fine utilization of rotation symmetry via the equivariant network, which can reduce the model parameters from 5,095,703 (InDuDoNet) to 4,723,309 (MEPNet). Besides, as observed from the bone marked by the green box, MEPNet alleviates the rotational-structure distortion generally existing in other baselines. This finely validates the effectiveness of embedding rotation equivariance.

\vspace{1mm}
\noindent{\bf Quantitative Evaluation.} 
Table~\ref{tabsyn} lists the average PSNR/SSIM on three testing sets.
It is easily concluded that with the increase of under-sampling rates, all these comparison methods present an obvious performance drop. Nevertheless, our MEPNet still maintains higher PSNR/SSIM scores on different testing sets, showing good superiority in generalization capability. Table~\ref{tab:160} reports the results on the DeepLesion-test with different sizes of metals under the ×4 sparse-view under-sampling rate. We can observe that MEPNet almost outperforms others, especially for the large metal setting, showing good generality.\footnote{More experimental results are included in {\itshape supplementary material}\label{exp}.}

\begin{table}[t]
\centering
\vspace{0mm}
\caption{Average PSNR (dB) and SSIM of different methods on three testing sets.}\vspace{-3mm}
% $^{\text{*}}$ means we adopt the pre-trained model released by the authors \cite{wei2019semi}.
%\begin{tabular}{@{}c|c@{}c|c|c|c|c|c|c@{}}
\tiny
\setlength{\tabcolsep}{5pt}
\begin{tabular}{l|ccc|ccc|ccc}
\Xhline{0.6pt}
\multirow{2}{*}{Methods}   & \multicolumn{3}{c|}{DeepLesion-test} & \multicolumn{3}{c|}{Pancreas-test}  & \multicolumn{3}{c}{CLINIC-test}\\
& ×8 & ×4 &×2 & ×8 & ×4 &×2  & ×8 & ×4 &×2\\
\Xhline{0.6pt}
\multirow{2}{*}{Input}             &12.65         &13.63        &16.55             &12.37          &13.29                  &16.04   &13.95        &14.99        &18.04           \\
 &0.3249         &0.3953        &0.5767  &0.3298         &0.3978           &0.5645             &0.3990          &0.4604            & 0.6085            \\
\hline
\multirow{2}{*}{FBPConvNet~\cite{jin2017deep}}   &25.91            &27.38            &29.11               &24.24       &25.44             &26.85    &27.92   &29.62   &31.56 \\
&0.8467            &0.8851            &0.9418               &0.8261         &0.8731           &0.9317 &0.8381 &0.8766 &0.9362        \\
\hline
\multirow{2}{*}{DuDoNet~\cite{lin2019dudonet}}         &34.33   &36.83                         &38.18         &30.54           &35.14                  &36.97  &34.47 &37.34    &38.81       \\
   &0.9479 &0.9634                &0.9685                     &0.9050         &0.9527           &0.9653  &0.9157 &0.9493   &0.9598      \\
\hline
\multirow{2}{*}{InDuDoNet~\cite{wang2021indudonet}}      &\underline{37.50}  &\underline{40.24}           &\underline{40.71}         &\bf{36.86}           &\underline{38.17}             &\underline{38.22}           &\underline{38.39}  &\underline{39.67}         &\underline{40.86}  \\
 &\underline{0.9664}            &\underline{0.9793}           &\underline{0.9890}         &\underline{0.9664}          &\underline{0.9734}             &\underline{0.9857}           &\underline{0.9572}   &\underline{0.9621}     &\underline{0.9811}     \\
\hline
\multirow{2}{*}{MEPNet (Ours)} &\bf{38.48} &\bf{41.43}                 &\bf{42.66}         &\underline{36.76}           &\bf{40.69}             &\bf{41.17}           &\bf{39.04}    &\bf{41.58}   &\bf{42.30}      \\
  &\bf{0.9767}           &\bf{0.9889}           &\bf{0.9910}                     &\bf{0.9726}         &\bf{0.9872}           &\bf{0.9896}  &\bf{0.9654} &\bf{0.9820} &\bf{0.9857}            \\
   \hline
\end{tabular}
\label{tabsyn}
\vspace{-1mm}
\end{table}
\begin{table}[t]
\centering
\caption{Average PSNR (dB)/SSIM of the comparing methods on DeepLesion-test with the ×4 under-sampling rate and different sizes of metallic implants.}\vspace{-3mm}
% $^{\text{*}}$ means we adopt the pre-trained model released by the authors \cite{wei2019semi}.
%\begin{tabular}{@{}c|c@{}c|c|c|c|c|c|c@{}}
\tiny
\setlength{\tabcolsep}{2.9pt}
\begin{tabular}{l|c|c|c|c|c|c}
   \hline
Methods    & \multicolumn{5}{c|}{ Large Metal \quad \quad   \quad\quad  $\longrightarrow$    \quad   \quad\quad \quad         Small Metal}                & Average      \\
   \hline
Input             &13.68/0.3438              &13.63/0.3736              &13.61/0.4046               &13.61/0.4304              &13.60/0.4240              &13.63/0.3953              \\
FBPConvNet~\cite{jin2017deep}              &26.15/0.7865              &26.96/0.8689              &27.77/0.9154              &27.98/0.9216              &28.03/0.9331              &27.38/0.8851\\
DuDoNet~\cite{lin2019dudonet}        &31.73/0.9519 &33.89/0.9599 &37.81/0.9667  & 40.19/0.9688 & 40.54/0.9696 & 36.83/0.9634 \\
InDuDoNet~\cite{wang2021indudonet}         & \underline{33.78}/\underline{0.9540} & \underline{38.15}/\underline{0.9746} & \underline{41.96}/\underline{0.9873}  & \underline{43.48}/\underline{0.9898} & {\bf{43.83}}/\underline{0.9910} & \underline{40.24}/\underline{0.9793} \\
MEPNet (Ours) & \textbf{37.51/0.9797} & \textbf{39.45}/\textbf{0.9879} & \textbf{42.78}/\textbf{0.9920} & \textbf{43.92}/\textbf{0.9924} & \underline{43.51}/\textbf{0.9924}& \textbf{41.31}/\textbf{0.9889} \\    \hline
\end{tabular}
\vspace{-5mm}
\label{tab:160}
\end{table}

\vspace{-4mm}
\section{Conclusion and Future Work}
\vspace{-2mm}
% In this paper, we proposed a joint spatial and Radon domain reconstruction model, namely InDuDoNet, for this metal artifact reduction (MAR) task and design an iterative optimization algorithm, which improves the interpretability of our framework. Extensive experiments were conducted on not only the synthesized, but also the clinical data. The experimental results demonstrated the effectiveness of our dual-domain MAR approach---a new state-of-the-art was achieved.
In this paper, for the SVMAR task, we have constructed a dual-domain reconstruction model and built an unrolling model-driven equivariant network, called MEPNet, with a clear working mechanism and strong generalization ability. These merits have been substantiated by extensive experiments. Our proposed method can be easily extended to more applications, including limited-angle and low-dose reconstruction tasks. A potential limitation is that consistent with~\cite{zhou2022dudodr,wang2021indudonet}, the data pairs are generated based on the commonly-adopted protocol, which would lead to a domain gap between simulation settings and real clinical scenarios. In the future, we will try to collect clinical data captured in the sparse-view metal-inserted scanning configuration to comprehensively evaluate our method. 

% By correspondingly unfolding every step involved in this algorithm into each network module, we have easily constructed the entire network architecture with fine interpretability. This has been visually verified by model visualization for easy understanding the working mechanism of our network. Based on synthesized and clinic datasets, comprehensive experiments on MAR task as well as downstream multi-class pelvic fracture segmentation task, have shown great superiority of our method as compared with current SOTA MAR approaches both visually and quantitatively.

%\newpage
\bibliographystyle{splncs04}
\bibliography{egbib}
\end{document}